\documentstyle[amsfonts,epsfig,12pt]{article}
\def\mypagenumber{1}

\def\myend{\end{document}}

\normalsize

\newcounter{sxn}

\newcounter{axn}

\date{}

\newdimen\mybaselineskip
\newcommand{\beeq}{\begin{equation}}
\newcommand{\eneq}{\end{equation}}
\newcommand{\be}{\begin{eqnarray}}
\newcommand{\ee}{\end{eqnarray}}
\newcommand{\bpic}{\begin{picture}}
\newcommand{\epic}{\end{picture}}

\def\dd{\partial}
\def\la{\raise.16ex\hbox{$\langle$} \, }
\def\ra{\, \raise.16ex\hbox{$\rangle$} }

\def\psibar{ \psi \kern-.65em\raise.6em\hbox{$-$} }
\def\mbar{ m \kern-.78em\raise.4em\hbox{$-$}\lower.4em\hbox{} }

\def\L{ {\cal L} }
\def\ep{\epsilon}

\def\n@space{\nulldelimiterspace=0pt \mathsurround=0pt }
\def\huge#1{{\hbox{$\left#1\vbox to 20.5pt{}\right.\n@space$}}}

\def\myskip{\noalign{\kern 8pt}}
\def\myeqspace{\noalign{\kern 10pt}}

\def\boxit#1{$\vcenter{\hrule\hbox{\vrule\kern3pt
    \vbox{\kern3pt\hbox{#1}\kern3pt}\kern3pt\vrule}\hrule}$}
\def\bigbox#1{$\vcenter{\hrule\hbox{\vrule\kern5pt
     \vbox{\kern5pt\hbox{#1}\kern5pt}\kern5pt\vrule}\hrule}$}

\def\ignore#1{{}}

\begin{document}

\bibliographystyle{unsrt}
\footskip 1.0cm

\thispagestyle{empty}
\setcounter{page}{\mypagenumber}

             
\begin{flushright}{OUTP-00-36-P\\
}

\end{flushright}

\vspace{2.5cm}
\begin{center}
{\LARGE \bf {Magnetic  Symmetries and Vortices In Chern-Simons Theories  }}\\
\vskip 0.5 cm   
\vspace{1.3cm}
{\large Gerald Dunne$^{a,b}$ \footnote{
dunne@phys.uconn.edu},\hskip 0.3 cm  Alex Kovner$^b$ \footnote{
kovner@thphys.ox.ac.uk},\hskip 0.3 cm 
Bayram Tekin$^b$ \footnote{
tekin@thphys.ox.ac.uk}
}\\
\vspace{.5cm}

\vspace{.5cm}
$^a${\it Department of Physics, University of Connecticut, Storrs CT 06269, USA}
\\
$^b${\it Theoretical Physics, University of Oxford, 1 Keble Road, Oxford,
OX1 3NP, UK}\\  
\end{center}

\vspace*{2.5cm}


\begin{abstract}
\baselineskip=18pt
We study the locality properties of the vortex operators in compact 
U(1) Maxwell-Chern-Simons  and SU(N) Yang-Mills-Chern-Simons 
theories in 2+1 dimensions. We find that these theories 
do admit local vortex operators and thus in the UV regularized versions
should contain stable magnetic vortices. In the continuum limit however
the energy of these vortex excitations generically 
is logarithmically UV divergent. Nevertheless the classical analysis
shows that at small values of CS coefficient $\kappa$
the vortices become light. It is conceivable that they in fact become
massless and condense due to quantum effects 
below some small $\kappa$. If this happens the magnetic symmetry  breaks
spontaneously and the theory is confining.

\end{abstract}
\vfill

PACS: ~   


 
\newpage



\normalsize
\baselineskip=22pt plus 1pt minus 1pt
\parindent=25pt
\section{Introduction.}

Recently there has been proposed \cite{kaks,kak} a very general argument (reviewed below)
connecting the realization of magnetic $Z_N$ symmetry, and the vacuum expectation values
of the spatial 't Hooft loop V and spatial Wilson loop W in SU(N) gauge theory. The 
motivation for these arguments is the study of confinement/deconfinement phase transitions
in such theories. In 2+1 dimensions the role of the spatial 't Hooft loop operator
is played by the magnetic vortex operator $V(x)$ which creates a magnetic vortex at $x$.
The magnetic vortex operator is a  canonical operator that acts on the physical
Hilbert space of theory, unlike the Polyakov loop which is also used to study the
phase structure of gauge theories, but which is not a canonical operator \cite{kaks,kak}.
An important part of the aforementioned argument concerns the {\it locality} of this
magnetic vortex operator. In this note we study the locality properties of the
magnetic vortex operator in 2+1 dimensional gauge theories when a Chern-Simons term is
included for the gauge field. We consider first the case of compact U(1) and then SU(N).
Including the Chern-Simons term does not alter the classical compact gauge symmetries of the
system, but we might expect new behavior because the Chern-Simons term generates massive
gauge degrees of freedom at the perturbative level \cite{Deser}.

Several related issues have been studied previously. In the absence of the Chern-Simons
term, compact QED in 2+1 dimensions is confining due to monopoles \cite{polyakov},
while SU(N) Yang-Mills in 2+1 dimensions is confining due to the condensation of $Z_N$
monopoles \cite{thooft}. The existence of similar monopoles with a Chern-Simons term 
included has been analyzed for compact $U(1)$ in \cite{affleck,fradkin,klee}, and for SU(N) in
\cite{dhoker,pisarski,tekin}. The general picture emerging from these studies is that
confinement of electric charge is destroyed by the inclusion of the Chern-Simons term
due to the binding of monopole-antimonopole pairs. However, these instanton-based 
approaches involve complex instantons, whose physical interpretation is acknowledged to
be incompletely understood \cite{tekin}. We thus feel it will be helpful to analyze these models
in a different but complementary manner. Compact Maxwell-Chern-Simons theories have also
been studied on the lattice in \cite{cristina}, also revealing bound monopole-antimonopole 
pairs. The strong coupling limit of SU(N) Yang-Mills-Chern-Simons theories, and in particular
the Polyakov loop, have been analyzed in \cite{grignani} using the connection to topological
field theories. And a Hamiltonian analysis \cite{nair}
of SU(N) Yang-Mills-Chern-Simons theories indicates that the Wilson loop is expected not to
have an area law behavior. Cornwall has argued \cite{cornwall} that there is a phase 
transition at a finite
value of the Chern-Simons coefficient due to an interplay of perturbative and non-perturbative
mass generation effects. As further motivation, much is known about the 
existence and properties of magnetic vortices in Chern-Simons theories coupled to matter fields 
\cite{kimyeong,dunne}. Here, instead, we want to focus on the question of vortices in the theory
with just gauge fields, and no additional matter fields.

The key objects in this discussion are the spatial Wilson loop operator (along the
spatial boundary of the system):
\be
W={\rm Tr}{\cal P}\exp\{i\oint_{C\rightarrow\infty} dx_iA_i\}
\ee
and the  magnetic vortex creation operator $V(x)$, whose precise definition will be given below
for the U(1) and SU(N) cases [see (\ref{vortex3}) and (\ref{v2})]. In the SU(N) case, W 
generates the discrete
$Z_N$ global symmetry, while $V(x)$ can be viewed as a canonical local order
parameter for this symmetry \cite{kaks,kak}. Together, W and V satisfy the 't Hooft
algebra \cite{thooft}:
\be
W\, V(x)\, W^\dagger = e^{\frac{2\pi i}{N}}\, V(x)
\label{th}
\ee
The general argument developed in \cite{kaks,kak} for SU(N) Yang-Mills, without
a Chern-Simons term, can be briefly summarized as follows.

First, one establishes that $V(x)$ is a local canonical order parameter that maps
physical states to physical states, and so can be used to distinguish the
phases of the theory in a gauge invariant manner. Given this, consider the theory
at zero temperature. The spatial Wilson loop operator (the generator of the magnetic
$Z_N$ symmetry) acts on the ground state by transforming the $Z_N$-non invariant fields.
There are two natural possibilities:

(i) the ground state is $Z_N$ invariant, so that $<V>=0$. The operator W only has
an effect near the boundary, so W has a perimeter law behavior. In this case,
we expect vortex states in the spectrum, to carry the unbroken $Z_N$ symmetry.

(ii) the ground state breaks $Z_N$, so that $<V>\neq 0$. Now W acts nontrivially
in the bulk so that W has an area law behavior. In this case we expect no
vortices in the spectrum.

At higher temperatures, these correspondences no longer hold. Even in an 
unbroken phase, the thermal ensemble involves states with nontrivial $Z_N$
charges, so that W acts significantly in the bulk and so generically has an area law
behavior even though the symmetry is restored.

In this note, we ask how much of this argument is modified by the inclusion of a Chern-Simons
term. We will show that the Wilson loop still generates the appropriate discrete symmetry.
But Yang-Mills-Chern-Simons (YMCS) theories are completely massive, and so the Wilson loop
is expected to have a perimeter law behavior \cite{nair}. By the previous argument it should 
then follow that the magnetic symmetry is unbroken and the spectrum should contain
vortex states. There is one way to avoid this conclusion.
The relation between the VEV of $W$ and $V$ only holds if $V$ is a local operator. 
The question of locality of $V$ has 
not been studied in YMCS theories. Thus it is possible that $V$ is nonlocal,
in which case the magnetic symmetry does not actually have a local order parameter. 
If that is the case the symmetry can be unbroken, 
$W$ may have perimeter law and there could still be
no magnetic vortex states. We note that a similar situation occurs
in YM theories with fundamental matter\cite{kovner3}.
This is the question we want to address in this note.
We want to examine more 
carefully the question of locality of the vortex operator in YMCS and 
understand whether the spectrum contains magnetic vortices.

In Section 2 we start our discussion by considering a simpler theory - compact
electrodynamics with CS term. Conceptual questions here are similar but the 
technical side is much simpler. Our results are somewhat surprising. We find
that generically the theory does indeed contain local vortex operators
and a global discreet magnetic symmetry which is unbroken.
Nevertheless in the continuum limit there are no magnetic vortices in 
the spectrum.
The reason is that the energy of such a vortex is logarithmically UV divergent. 
We find however that with a particular scaling of the CS coefficient (logarithmically
vanishing when UV cutoff is removed) the energy of the vortex becomes finite. This
suggests that the theory may indeed have a phase with finite energy vortex states
and vanishing photon mass. 
In Section 3 we extend our discussion to non-Abelian YMCS theories. We find
that here again the local gauge invariant vortex field exists.
The situation in the continuum limit appears to be 
similar to the Abelian case
with a possible vortex phase at small values of the CS coefficient.
Section 4 contains a short summary of our results.

\section{Compact QED with the Chern Simons term.}
The Lagrangian of Abelian Chern Simons theory in the formal continuum limit is
\be 
 \L =  -{1\over 4 g^2 } F_{\mu\nu} F^{\mu\nu}
 +{\kappa\over 2} \ep^{\mu\nu\rho} A_\mu \dd_\nu A_\rho
\label{Lagrangian1}
\ee
The gauge coupling 
$g^2$ has dimension of mass and $\kappa$ is dimensionless. Equations of motion
read as $\dd_{\nu}F^{\mu\nu}=  \kappa g^2 \ep^{\mu\nu\rho} \dd_\nu A_\rho$.
The mass of the gauge particle is $M = \kappa g^2$. 
The canonical structure of the theory is simplest in the Hamiltonian gauge, $A_0= 0$.

The Hamiltonian is
\be
H= {1\over 2g^2}( E_i^2 +B^2)
\label{hamiltonian}
\ee 
with canonical momenta related to the time derivatives of the fields by
$\Pi^i= -{1\over g^2} {\dot{A}}_i + {\kappa\over 2}\epsilon^{ij}A_j$. 

The gauge fields and canonical momenta form the canonical algebra, and
the algebra involving the electric fields is
\be
[E_i(\vec{x}),E_j(\vec{y})] = -i \kappa g^4\epsilon_{ij}
\delta^2(\vec{x}- \vec{y}), \hskip 1 cm 
[A_i(\vec{x}),E_j(\vec{y})] = -ig^2 \delta_{ij} \delta^2(\vec{x}- \vec{y})
\ee
The Gauss law,  
\be
 \dd_iE^i -\kappa g^2 \epsilon^{ij}\dd_i A_j= 0
\ee
 on a given
spatial slice $\Sigma$ which we take to be the plane, 
generates time independent local gauge transformations. The elements of the
local gauge group near a point $x_0\in \Sigma$ take the form 
\be
U(x_0) = \mbox{exp} \big\{{1\over g^2} i\int_{\Sigma} d^2x\,\lambda(x,x_0)
\big( \dd_iE^i - g^2\kappa \epsilon^{ij}\dd_i A_j \big)\big \}    
\label{gausslaw}
\ee
such that $UAU^{-1} = A + d\lambda$. In the non-compact theory 
$\lambda \in {\mathbb{R}}$ 
should be a single-valued function on $\Sigma$ such that the eigenvalue
of the operator (\ref{gausslaw}) on physical states is unity.
Singular $\lambda$'s correspond to transformations which
are in general nontrivial on the physical states. 

Our interest however is in the compact theory. This means that
magnetic vortices of flux $2\pi$ must be physically unobservable.
As discussed in \cite{Kovner} this amounts to
further restricting the physical Hilbert space to states
which are trivial under the action of the vortex operator.

In other words
certain large gauge transformations must act on the physical states
trivially in the compact theory as opposed to the non-compact one.  The
compact gauge group therefore includes these 
singular gauge transformations in addition to the regular ones, 
which form the gauge group in the noncompact theory.
Consider
a multi-valued angle function 
$\theta(x,x_0)$ which is singular at one point and
has a discontinuity along a straight 
curve $C(x_0)$ that starts at the point $x_0$
and goes to infinity. The operator of the gauge transformation with this
singular gauge function creates a magnetic vortex.
Its explicit form (after partial integration and dropping a boundary term 
owing to the
fact that all gauge invariant fields decay at infinity) is
\be
\tilde V(x_0) = 
\mbox{exp} \{ -{1\over g^2} i\int_{\Sigma} d^2x\,{\tilde{\dd_i}} 
\theta(x - x_0) \Bigg( E^i -g^2\kappa \epsilon^{ij} A_j \Bigg)\}    
\label{vortex}
\ee         
We have defined   
\be
{\tilde{\dd_i}}\theta(x - x_0) = \dd_i\theta(x - x_0) - 
2\pi\epsilon_{ij} c(x)_j \delta (x- C(x_0) ) =
 {\epsilon_{ij}(x- x_0)_j\over (x-x_0)^2} 
\ee
$c(x)_j$ is a unit vector tangent to the curve $C(x,x_0)$.

We may want to include $\tilde V$ into the compact gauge group.
However to be part of the gauge group, it must commute (at least weakly)
with other elements of the group.
One can check explicitly that  $\tilde V(x,x_0)$ does not commute with the elements
of the noncompact group.
To rectify this situation we define, following \cite{Kovner} 
a slightly modified operator
\be
V(C, x_0)= \mbox{exp}{2\pi i\over g^2}\epsilon_{ij}\int d^2 x c(x)_j 
\delta (x- C(x,x_0))E_i(x)  
\ee
This operator is  merely a ``collection''  of the
electric fields which are perpendicular to the curve $C(x_0)$.  
More explicitly one can write it in the following form
\be
V(C, x_0)= \mbox{exp}{2\pi i\over g^2}\epsilon_{ij}\int_C dl_i\,E_j(x)  
\label{vortex3}
\ee
Gauge invariance of this operator, $[V, U]=0$, follows immediately.
We also need to check the commutativity of $V(x)$ with $V(y)$.
Defining the volume form $ v = {1\over 2} \epsilon_{ij} dx^i\wedge dx^j$ 
on $\Sigma$ we have the following commutation rules,
\be
V(C_0)V(C_1)= V(C_1) V(C_0){\mbox{exp}}
\Bigg \{i 8 \pi^2 \kappa L(C_0,C_1) \Bigg \}    
\ee
Where $L(C_1,C_2)= \int_\Sigma v \delta^2 (C_0- C_1)$. It is clear that
if the curves cross each other then $L= \pm 1$ and 
if they are parallel $ L= 0$.
In order that $V$ be a Lorentz scalar the commutator should not depend on
the curves $C_1$ and $C_2$
To guarantee this
we need to set 
\be
4\pi\kappa = k \in {\mathbb{Z}}
\ee
We find therefore that the requirement of compactness quantizes 
the coefficient of the Chern Simons term very much like
in the non-Abelian theory .

To see that (\ref{vortex3})
creates magnetic vortices of integer strength it is enough to 
consider the commutator
\be
[B(x), V^m(x_0)]= 2\pi m \delta^2(x- x_0)V^m(x_0), \hskip 1 cm  
m \in {\mathbb{Z}}.
\ee
Therefore $V^m(x_0)$ creates a point-like magnetic vortex of vorticity 
$m$ and magnetic flux $2\pi m$. 
Being gauge invariant this operator
also creates an electric charge
\be
Q= {1\over g^2} \int_\Sigma d^2 x \dd_i E^i 
= {k\over 4\pi} \int_\Sigma d^2 x B = {mk\over 2}
\ee
Since $V$ has to be included in the gauge group, the magnetic flux and the 
electric charge created by it must be unobservable.
Therefore the Hamiltonian of the theory must commute with $V$. 
The noncompact Hamiltonian eq.(\ref{hamiltonian}) does not quite do the job.
It should be modified but in such a way that in the continuum limit the same form is
recovered for smooth fields.
The modified Hamiltonian that satisfies these conditions has been 
suggested in \cite{Kovner}.
Since the UV structure is important for our considerations, it
is most usefully presented in the lattice notations
\be
H_B= {1\over a^4 g^2 n^2}\sum_x\big( 1- \mbox{Re}\,e^{i n a^2 B(x) } \big), 
\hskip 1 cm H_E= {m^2 g^2\over 4\pi^2 a^2 }\sum_x
\big( 1- \mbox{Re}\,e^{i2\pi {a \over m g^2 }\epsilon_{ij}\hat{n}_j E_i(x)} \big), 
\label{latticehamiltonian}
\ee
$a$ is the lattice spacing and $ m,n \in {\mathbb{Z}}$ and  $\hat{n}_j$ is
the unit vector parallel to the link. 
The normalization of the electric and magnetic terms is such 
that in the naive continuum limit $a\rightarrow 0$ they reduce to $B^2$ and $E^2$
respectively.

Using the Gauss' law one can see 
that if $2n = k$ , the magnetic part
$H_B$ becomes a combination of the vortex operators $V$. 
Therefore without loss of generality we assume $2n < k $. 
For $m=1$ the electric part of the Hamiltonian is also a sum of a fundamental
vortex and anti-vortex, and we take $m > 1$.

Now that we have the formulation of the compact CS QED we can ask about the 
locality properties of
vortex operators. The operator $V$ we have considered so far is of no interest of 
itself, since is is trivial on all physical states. We thus have to look at
the operators which create magnetic flux smaller than $2\pi$
\be
V_p(C, x_0)= \mbox{exp}{2 p \pi i\over g^2}\epsilon_{ij}\int_C dl_i\,E_j(x)  
\label{nonintegervortex},\hskip 1 cm p \in {\mathbb{Q}}
\ee  
Here $p$ is a rational number $p \in (0,1)$. 
The question we are asking is, are there such values of $p$ for which
$V_p$ is a gauge invariant local operator.
The gauge invariance with respect to the
noncompact gauge group is straightforward, since $V_p$ only depends on the
electric field, and the electric field itself is gauge invariant.
However $V_p$ should also commute with the "fundamental" vortex $V$, since $V$ is
part of the gauge group.
Therefore we have
\be
[V,V_p]=0 \hskip 1 cm \Longrightarrow  k p = l \in  {\mathbb{Z}} \hskip 1 cm
\mbox{and}\hskip 0.3 cm l < k.   
\label{condition1}
\ee
This condition is already informative. For example
we see that there
are no non-trivial vortices in $k=1$ theory. 
The condition of locality requires that $V_p(x)$  commute with each other
at different points $x$ and $y$. This commutator also should be independent of the
contour $C$ in the definition eq.(\ref{nonintegervortex}).
\be
[V_p(x),V_p(y)]=0 \hskip 1 cm \Longrightarrow  
k p^2 = r \in  {\mathbb{Z}} \hskip 1 cm  
\label{condition2}
\ee
Both equations (\ref{condition1}) and (\ref{condition2}) have to be satisfied
for the
existence of non-trivial local vortices. 
Whether it is possible or not to satisfy these equations clearly depends on the
CS coefficient $k$.
For example  
there are no solutions for $k= 2$ and $k= 3$ theories. 
For $k=4$ we can choose $l= 2$ and this gives a vortex of 
vorticity $p = 1/2$. In general one can solve the constraints in the following
way. Writing CS coefficient in terms of its prime factors, $k= q_1\,q_2\,q_3...
\,\,q_m$, where all $q_i$ are not necessarily different,
one has the following two conditions to satisfy
\be
q_1\,q_2\,q_3...\,q_m\, p = l, \hskip 1 cm 
q_1\,q_2\,q_3...\,q_m\, p^2 = r. 
\ee  
The first condition is solved if $p$ divides $k$ which means, without
loss of generality, we have
\be
p =  {1\over q_1\,q_2\, q_3...\,q_i}, \hskip 0.5 cm \mbox{where} \hskip 0.5 cm
i< m  
\ee
Using this in the second condition one can see that the most general form of
$k$ which allows vortices will be
\be
&&\mbox{If}\hskip 0.3 cm  k =  t^2 z  \hskip 0.5 cm 
\Longrightarrow  p = {1\over t},\hskip 0.5 cm  
t \geq 2 \hskip 0.3 cm \mbox{and} \hskip 0.3 cm t,z \in   {\mathbb{Z}}    
\label{tz}
\ee
For example if $k$ is a prime number there are no solutions.
Generically it is easier to find a solution at large values of $k$.
 
The above relations also
show that should a solution exist, there is always a vortex of minimal vorticity. 
All the other local vortices are simple powers of this minimal vortex.
For example for $k = 36$,
the above conditions give three solutions (and their integer multiples) , $ p=(1/2, 1/3,1/6 )$.
Obviously ``p= 1/6'' is the minimal vortex. 
The minimal value of $p={1\over w}$ determines the global magnetic symmetry group
of the theory as $Z_{w}$.

One last requirement that $V_p$ should satisfy, is locality with 
respect to the energy density eq.(\ref{latticehamiltonian}). In obvious notation
\be
[h_E(x), V_p(y)]=0, \ \ \ \ x\ne y \Longrightarrow {kp\over m} \in   {\mathbb{Z}}
\ee
This can always be satisfied by choosing $m = k$.
To satisfy the other
 condition 
\be
[h_B(x), V_p(y)]=0  \Longrightarrow {n p\over 2} \in   {\mathbb{Z}}
\ee
we can take $n=2$. Certainly one can define other Hamiltonians which
will be compatible with the above conditions.
We see therefore that for those values of $k$ for which vortex operators 
are local with respect
to each other we can always choose the Hamiltonian such that they are 
also local relative to the Hamiltonian density.

Thus we conclude that for many values of $k$ local physical vortex operators
exist. 
They are order parameters for a global $Z_w$ magnetic symmetry. The value of $w$
is determined by $k$ through the solution of the equations for minimal $p$. 
Thus the argument described 
in the introduction applies and, at least in the lattice theory 
there are vortex states. Calculating their energy in the lattice theory
is not a simple matter. However 
the interesting question is whether these states survive in the 
continuum limit. That is to say, whether their energy
stays finite as the lattice spacing approaches zero.

We note that the continuum limit of the theory (\ref{latticehamiltonian})
is a slightly delicate matter. One must certainly take the limit
$a\rightarrow
0$.
However since we expect in the continuum the scaling $B$ and $E$ 
to be $ga^{-3/2}$,
to guarantee the expandability of the electric exponential in 
eq.(\ref{latticehamiltonian}) we should take
$gka^{1/2}\gg 1$. For the expandability of the magnetic exponential
we should keep $ga^{1/2}\ll 1$. These two conditions are compatible 
if $k\gg 1$. 
With this scaling the mass of the photon can be also kept
finite in the continuum limit.
One should keep in mind that 
this scaling of the couplings is sufficient to get the continuum limit, 
but it may not be
necessary. In particular it could certainly happen that at finite $k$ 
the scaling of $B$
and $E$ changes close to the cutoff and the continuum limit still exists.

In the continuum limit for smooth configurations of the fields
the theory is described by the Lagrangian eq.(\ref{Lagrangian1}).
However while solving continuum equations we may sometimes encounter
field configurations with fast variations. For these configurations it is important
to take into account the compactness of the theory. 
In particular consider the electric field created by the 
"minimal" vortex operator
$V_{1/w}$.
\be
[V(x),E_i(y)]=E_i(y)+e_i(x,y),\ \ \ \ e_i(x,y)=
{1\over w} g^4\kappa\hat n(y)_i \delta (y- C(x,y))
\label{field}
\ee
where $\hat n(y)$ is the vector tangential the curve $C$ at the point $y$.
Since the operator $V(x)$ is local,
its only observable action in the compact theory is at the point $x$. 
However if we just calculate the energy using the naive Hamiltonian 
eq.(\ref{hamiltonian})
we find infrared divergence proportional to the length of the curve $C$.
Clearly if faced with this type of configurations in continuum calculations
we should subtract this infrared divergence by hand. 
Rather than do this we find it convenient to think about it in the
following way. Let us split the general electric field configuration
into a smooth piece and a piece that contains arbitrary number
of strings of the type of eq.(\ref{field})
\be
E^i=E^i_{smooth}+e^i
\ee
and subtract the contribution of $e_i$ in the Hamiltonian.
The only remnant of $e_i$ then is in the Gauss' law, since $e_i$ of
eq.(\ref{field}) corresponds to a pointlike charge $ {k\over 2w}$ at the
point $x$. Thus the smooth field 
$E^i_{smooth}$ satisfies not the naive Gauss' law, but rather a modified one
\be
\dd_iE^i_{smooth}|_{{\rm mod}{k\over 2w}\delta^2(x)}   
-\kappa g^2 \epsilon^{ij}\dd_i A_j= 0
\label{gaussviolation}
\ee
In other words we can work entirely in terms of $E_{smooth}$ if we remember that
we may allow Gauss' law to be violated by the presence of 
pointlike charges of charge $\kappa g^2/w$. The appearance of $w$ in this way is 
the only remnant of the compactness of the theory. In the following we will
work in terms of the smooth fields but will drop the subscript $smooth$ for
brevity.

With this caveat in mind,
to determine the energy of the magnetic vortex in the continuum limit
we now should solve the continuum equations of motion.
For a minimum vorticity solution $1/w$ 
following \cite{Nielsen, Khare} one can take the time independent symmetric 
ansatz,
\be
A_i(r)=  \epsilon_{ij}{x_j\over r^2}[ g(r) - {1\over w}] \hskip 1 cm 
A_0(r)=  h(r) 
\ee
The equations of motion read 
\be
g''(r) -{1\over r} g'(r) -rM h'(r) =0, \\
h''(r) +{1\over r} h'(r) -{M\over r} g'(r) = 0,
\label{gausslaw2}
\ee
where $ M= g^2\kappa$. We are looking for the solutions with
vorticity $1/w$. The magnetic field is $B = - {1\over r} g'(r)$ so we impose
$g(0)= 1/w$ and  $g(\infty)= 0$ and we also demand that the fields decay
exponentially at infinity. The solution under these conditions 
can be found as
\be
&&g(r)= {Mr\over w}K_1(Mr), \hskip 1 cm B(r)= {M^2\over w} K_0(Mr) \\
&& h(r)= -{M\over w} K_0(Mr)\hskip 1cm 
E^i= -{x^i  M^2\over wr}\,K_1(Mr), 
\label{solution}
\ee
where $K_0, K_1$ are the Bessel functions.The energy of this vortex follows as
\be
{\cal{E}}= {\pi\over g^2} \int_0^\infty r\,dr \{ ({dh\over dr})^2 + 
{1\over r^2} ({dg\over dr})^2 \}  
\ee     

\be
{\cal{E}} = {\pi\kappa\over w^2} M \big( -\gamma_E +
\mbox{ln}2 - \mbox{ln}{M\over \Lambda} \big)  
\label{energy}
\ee
where $\Lambda$ is the ultraviolet cutoff scale.

This result warrants several comments. First, we see that the energy 
of the vortex is
IR finite. This is closely related to the fact that the locality of the 
operator $V_{1/w}$ allowed us to "violate" the Gauss' law.
Looking at the electric and magnetic fields in eqn (\ref{solution}) 
we indeed see that the naive Gauss' law is violated precisely by the amount
allowed by eq.(\ref{gaussviolation}). Without this deficit in the Gauss law any 
solution would have infrared divergent energy. This is because
any nonvanishing magnetic flux $\Phi$ would require a long range
electric field $E_i\propto g^2 \Phi \kappa{x_i\over x^2}$ to satisfy the Gauss' law.
The Coulomb energy of this field is logarithmically divergent in IR. The absence
of this long range piece in the field and the associated IR divergence in 
the energy is what distinguishes the compact and the noncompact theory.

Second, the energy is UV divergent. Thus the magnetic vortices do not 
survive in the 
continuum limit as finite energy excitations. This of course does not contradict
our original argument. 
The $Z_{w}$ symmetry is unbroken in the vacuum and excitations carrying the
quantum numbers of this symmetry are very heavy. 
The symmetry therefore seems 
completely irrelevant for the low energy dynamics.
The curious thing though is that although the energy of the vortex is large, 
it is still
much lower than the natural ultraviolet scale $\Lambda$. The vortices therefore
are not genuine ultraviolet objects in the lattice theory, but rather occupy 
an intermediate scale between the UV scale $\Lambda$ and the IR scale M.
In fact bringing $M$ down to zero makes vortices light.
At $M=\mu\ln^{-1}{\Lambda\over \mu}$
the energy of the vortex is finite\footnote{The reader may wonder why we 
are not bothered
by the factor $\kappa$ in eq.(\ref{energy}). After all we saw 
that in the continuum limit naturally $\kappa\rightarrow\infty$. The point is that
the flux of the minimal vortex generically scales as $w^2\propto k $. For
example if $k=x^2$ with some integer $x$, then clearly the minimal
solution of the eq.(\ref{tz}) corresponds to $w=x$. Thus $k /w^2$ is 
finite in the continuum limit.}.
This behavior in fact is very reminiscent of vortices in the
Higgs phase of the Abelian Higgs model. 
The mass of such a vortex at weak coupling is
very large; $M\propto M_v^2/g^2\ln M_H/M_v$, where $M_v$ is the mass of the massive
photon and $M_H$ is the mass of the Higgs particle.
However as the photon mass decreases, that is as the
theory approaches the phase transition line, vortices become light.
On the phase transition itself they in fact become massless and condense 
in the Coulomb phase.
It is not unlikely that similar phenomenon occurs in our model. As $\kappa$
decreases at 
fixed $g^2$ the photon becomes lighter, and the mass of the vortex
also decreases. It could happen that at some value of $\kappa$ the vortices
actually become massless and drive a phase transition into a phase with broken
magnetic symmetry. A transition of
precisely such type was conjectured to happen in the lattice model 
at $ k =8$ in \cite{Kogan}
and was seen in the variational calculation of \cite{Kovner}.
Of course, within the naive continuum limit we consider here
we are unable to see such behavior. However the fact that the
vortices
 become
light within the validity of the naive continuum limit is quite suggestive 
in this respect.
It is interesting to note that if the vortices indeed condense, the
magnetic $Z_w$ symmetry is spontaneously broken. By virtue of the
argument given in the introduction this means that the low $k$ phase
is confining.

\section{Non-Abelian theories.}

We now want to extend our analysis to non-Abelian theories.
We will study the $SU(N)$ gauge theory with the 
the Lagrangian
\be
\L = {1\over 2g^2}\mbox{tr}F_{\mu \nu}F^{\mu 
\nu} -\kappa\epsilon^{\mu \nu
\lambda}\mbox{tr}\left(A_\mu \dd_\nu A_\lambda +{2\over 3} A_\mu
A_\nu A_\lambda \right) 
\ee
CS coefficient has the well known quantization, $4\pi\kappa= k$ \cite{Deser}. 
We denote 
$A_\mu = A_\mu^a T^a$ and $\mbox{tr}[T^a,T^b]= -{1\over 2}\delta^{ab}$  
The Lie algebra generators obey $[T^a, T^b]= f^{abc}T^c$.
The field strength and the covariant derivative are 
$F_{\mu\nu} = \dd_\mu A_\nu  -\dd_\nu A_\mu +[A_\mu, A_\nu]$, 
$D_\mu= \dd_{\mu} +[A_{\mu},\,]$. 
The classical equations of motion follow as
\be
D_\mu F^{\mu\nu} -{1\over 2}g^2 \kappa \epsilon^{\nu\lambda\mu}F_{\lambda
\mu} = 0
\ee
The canonical structure of this theory is similar to the Abelian case.
In the $A_0^a= 0$ gauge we have \cite{dunne}
\be
\Pi^a_i= -{1\over g^2} E_i^a + {\kappa \over 2}\epsilon^{ij}A_j^a,
\hskip 1 cm \mbox{where} \hskip 1 cm    E_i^a = \dot{A}_i^a.
\ee
The Hamiltonian is
\be
H= {1\over 2g^2}\{ (E^a_i)^2 +(B^a)^2 \}, 
\hskip 1 cm \mbox{where} \hskip 1 cm    
B^a =  {1\over 2}\epsilon_{ij}\, F^a_{ij}.
\label{hamiltonian2}
\ee 
The canonical algebra is
\be
[E_i^a(\vec{x}),E_j^b(\vec{y})] = -i\delta^{ab} \kappa g^4\epsilon_{ij}
\delta^2(\vec{x}- \vec{y}), \hskip 1 cm 
[A_i^a(\vec{x}),E_j^b(\vec{y})] = -ig^2\delta^{ab} \delta_{ij} 
\delta^2(\vec{x}- \vec{y})
\ee
In terms of the momenta the Gauss law is
\be
 (D_i\Pi_i)^a = - {\kappa\over 2} \epsilon^{ij}\dd_i A_j^a
\ee

In the non Abelian YMCS theory, the large Wilson loop still 
commutes with the Hamiltonian. This is obvious in the Hamiltonian formalism, 
since the commutation relation between the vector potential and the
chromoelectric field
is unaffected by the presence of the CS term. The form of the Hamiltonian
in terms of $A_i$ and $E_i$ is also the same as without the CS term.
Since $W$ is a function of $A_i$ only, its commutator
with the Hamiltonian is exactly the same as in the theory without CS. 
Therefore the fundamental Wilson loop still generates a symmetry.

Again our question is whether the theory admits local vortex operators.
In the non-Abelian theory our choices are more limited than in compact
QED. In the SU(N) Yang Mills theory, the only candidates for local operators
are those that create quantized flux \cite{thooft,kovner1}. The vortex operator
in YM theory is \cite{kovner1,kovner2}:
\begin{equation}
V(x)=\exp\{{ 4\pi i\over g^2N} \int_C dy^i
\epsilon_{ij}{\rm Tr}(YE_i(y))
\label{v2}
\end{equation}
where the hyper charge generator $Y$ is defined as
\begin{equation}
Y={\rm diag} \left(1,1,...,-(N-1)\right)
\end{equation}
and the electric field is taken in the matrix notation $E_i=T^aE^a_i$.
It can be proven that in SU(N) YM theory, this operator despite its nonlocal and 
gauge non-invariant appearance is in fact a local, gauge invariant, Lorentz scalar field 
\cite{kovner1,kovner2}. The way it was constructed there was to require that it satisfies 
the 'tHooft algebra \cite{thooft} with the fundamental Wilson loop
\be
V^\dagger(x)W(C)V(x)=\exp\{{ 2\pi i\over N}n(x,C)\}W(C)
\label{th2}
\ee 
with $n(x,C)$ being the linking number on the plane between the point $x$ 
and the closed curve $C$.

We claim that the operator in (\ref{v2}) is also the appropriate vortex operator
when a Chern-Simons term is included for the gauge field.
The commutation relation (\ref{th2}) is still satisfied by the expression (\ref{v2}).
However an additional requirement was that $V$ be gauge invariant.
Here we should be more specific what we mean by that. The expression (\ref{v2})
is not explicitly gauge invariant since it depends on the chromoelectric field in
the hyper charge direction. However for $\kappa=0$ it has been proven 
that the matrix elements
of $V$  between physical states (those that satisfy Gauss' law) and 
non-physical states (non-singlet under gauge transformations) vanish.
This means that when we calculate matrix elements of any number of operators $V$
between gauge invariant states, only intermediate 
states from the physical sector contribute and so the form (\ref{v2}) can be
safely used even though it is not explicitly gauge invariant.
Let us briefly recap the proof\cite{kak}.
The wave functional of any physical state in a theory without
Chern Simons term depends only on gauge invariant characteristics
of the vector potential, i.e. only on the values of Wilson loops over all possible 
contours. 
\begin{equation}
\Psi[A_i]=\Psi[\{W(C)\}]
\end{equation}
Consider the action on such a state of the operator $V(x)$ and its gauge transform
$V_\Omega(x)$.
\begin{eqnarray}
V|\Psi>&=&\Psi_V[A_i]=\Psi[\{VW(C)V^\dagger\}] \nonumber \\
V_\Omega|\Psi>&=&\Psi_V^\Omega[A_i]=\Psi[\{V_\Omega W(C)V^\dagger_\Omega\}]
\end{eqnarray}
The action of $V(x)$ and $V_\Omega(x)$ on 
the Wilson loop is 
identical - they both multiply it by the center group phase  
(which stays unaffected by $\Omega$) if $x$ is inside $C$ and do 
nothing otherwise.
Therefore 
\begin{equation}
V|\Psi>=V_\Omega|\Psi>
\label{prev}
\end{equation} 
for any physical state $\Psi$.
Thus we have
\begin{equation}
\Omega V|\Psi>= \Omega V\Omega^\dagger|\Psi>=
V|\Psi>
\end{equation}
where the first equality follows from the fact that a physical state is 
invariant under
action of any gauge transformation $\Omega$ and the second equality 
follows from eq.(\ref{prev}).
But this equation is nothing but the statement that the state
$V|\Psi>$ is physical, i.e. invariant under any nonsingular gauge
transformation. 
Thus we have proved that $V$ transforms a physical state into 
another physical state.

In the Chern-Simons theory the vortex operator should also 
be gauge invariant.
We thus have to check that it transforms a physical state into another 
physical state.
The difference with the pure YM theory is that the wave function of a 
gauge invariant state does not depend only on the Wilson loops.
The physical wave function should satisfy the following equation
\be
i(D_i{\delta\over\delta  A_i})^a \Psi[A]=
{\kappa\over 2}\epsilon_{ij}\dd_i A_j^a \Psi[A]
\ee
The general form of $\Psi$ has been determined in \cite{nair} in terms
of certain nonlinear variables. For our purposes we find it more
convenient to work directly in terms of the vector potentials $A_i$.
Let us take $\Psi$ in the form
\be
\Psi = \exp\{- i S\}
\ee
Then the eikonal $S$ satisfies a linear inhomogeneous equation
\be
D_i^{ab}{\delta\over\delta  A_i^b}S[A]=
{\kappa\over 2}\epsilon_{ij}\dd_i A_j^a
\label{ds}
\ee
The solution of the homogeneous equation is indeed
any functional that depends on Wilson loops $S_0[W]$.
We can find a particular solution of the inhomogeneous equation
using the following argument.    
$S[A]$ must be a functional 
whose change under a standard
gauge transformation of vector potentials $\delta A_i=D_i\lambda$
is proportional to 
${\kappa\over 2} \int d^2x \epsilon_{ij}\dd_i A_j^a\lambda^a $.
Such a functional can be represented as a Chern Simons action on a space
with a boundary. Let us introduce an additional coordinate $\tau\in [-\infty,1]$
and functions of three coordinates $A_i(x,\tau)$ so that at the boundary $\tau=1$,
the value of these functions is equal to the value of the vector potentials in 
our theory $A_i(x,\tau=1)=A_i(x)$. Let us write the Chern Simons term 
(in the Weyl gauge) on this manifold
\be
S_{CS}=\int_{-\infty}^1d\tau\int d^2x \epsilon_{ij}A_i^a(x,\tau)\dot A_j^a(x,\tau)
\label{cs}
\ee
Under the $\tau$ independent gauge transformation this action changes by a 
boundary term
\be
\delta S_{CS}=-\int d^2x\,\epsilon_{ij}\lambda^a(x)\dd_iA^a_j(x,\tau=1)=
-\int d^2x\,\epsilon_{ij}\lambda^a(x)\dd_iA^a_j(x)
\ee
which is precisely of the form required to satisfy eq.(\ref{ds}).
A particular solution of eq.(\ref{ds})
is therefore
\be
S_p= -{\kappa\over 2} S_{CS}= -{\kappa\over 2}\int_{-\infty}^1 d\tau\int d^2x 
\epsilon_{ij}A_i^a(x,\tau)\dot{A}_j^a(x,\tau)
\ee
The introduction of the extra coordinate $\tau$ and the expression eq.(\ref{cs})
is not at all unnatural. One should view this extra coordinate as parameterizing
a curve in the field space. With this interpretation we have
\be
d\tau\dot A_j(x,\tau)=\delta A_i
\ee
and
\be
\int_{-\infty}^1d\tau\int d^2x \epsilon_{ij}A^i(x,\tau)\dot A_j(x,\tau)=
\int_{\cal C} \delta A_i\epsilon_{ij}A_j
\ee
where the line integral is taken over the trajectory ${\cal C}$
in the field space which ends at the point $\{A_i(x)\}$.

We have thus determined the general form of the wave function 
of a physical state in the YMCS theory to be
\be
\Psi[A]=\exp\{ i{\kappa\over 2}\int_{-\infty}^1d\tau\int d^2x 
\epsilon_{ij}A^i(x,\tau)\dot A_j(x,\tau)\}\Psi_0[W]
\ee
Now it is straightforward to
see how the vortex operator acts on it.
Under the action of the vortex operator
\be
V(x)\,A_i^a(y,\tau)\,V^{\dagger}(x) =  A_i(x,\tau) +
{4\pi\over N}\mbox{Tr}YT^a  \epsilon_{ij}\int dz_j \delta^2 (z- C(x,y))
\ee
Remembering that $E_i^a= -g^2 \Pi_i^a + g^2 \kappa\epsilon_{ij}A_j^a$, 
we see that
the change in the phase factor in the wave functional is exactly cancelled by 
the $A$-dependent term in the vortex operator eq.(\ref{v2}).
\be
V(x)\,e^{-i S_p}\,V^{\dagger}(x)= e^{-i S_p}.
\ee
Thus
\be
V\Psi[A]
=\exp\{i{\kappa\over 2} \int_{-\infty}^1d\tau\int d^2x 
\epsilon_{ij}A^i(x,\tau)\dot A_j(x,\tau)\}\Psi_0[V^\dagger WV]
\ee
Clearly a gauge transformed vortex operator $V_\Omega$ has exactly the 
same action on the
wave functional $\Psi[A]$, 
\be
V\Psi=V_\Omega\Psi
\ee
which establishes gauge invariance of $V$ in the same sense as in the 
YM theory. 

We now can check the locality of the operator $V$ by calculating
straightforwardly the relevant commutation relation.
A simple calculation gives  $[V(x), V(y)]= 0$.
Thus the operators are local with respect to each other. When
considering the locality with respect to the Hamiltonian density we
are faced with the same ambiguity as in the Abelian theory. The 
electric part of naive
continuum Hamiltonian is not local relative to $V$, since $E_i$ is
shifted by the action of $V$ along the curve $C$. Just like in
the Abelian case one should consider a properly regularized version of
$H$ in order to be able to draw a definite conclusion. In the
non-Abelian case such a regularized Hamiltonian is not
available. However in the Abelian case we saw that there is quite a
lot of flexibility in defining such a regularized version. In
particular we saw that whenever the vortex operators were local with
respect to each other, we were always able to define the local Hamiltonian
density. We expect that this situation persists
in the non-Abelian theory too.

The situation in the continuum limit is again similar to the Abelian
case.
There are no finite energy solutions of the non-Abelian equations of
motion which have finite vorticity. The only way to find such
solutions would be again to relax the Gauss' law constraint by
allowing point like charges which correspond to the singular
chromoelectric 
field created by $V$. However again those IR finite configurations
will have UV logarithmically divergent energy.
In fact taking Abelian ansatz the YM equations of motion reduce to
those we considered in the previous section and thus lead to the same
energy dependence on the UV cutoff.
Strictly speaking this conclusion is only valid for large enough value
of $k$, since for small $k$ quantum corrections to this classical
analysis may be large. Thus again it is possible that at small $k$ the
theory is in a different phase as suggested in \cite{cornwall}.

\section{Summary.}
To summarize, we have studied the question of locality of the vortex
creation operator in compact Chern- Simons theories. 
We have found that compact CS QED does admit local vortex operators
for many values of the CS coefficient $k$. The energy of the vortex
excitations however generically is 
logarithmically UV divergent in the continuum
limit. With a particular scaling of the CS coefficient these
vortices become light and might condense at small values of $k$.
Our results for non-Abelian CSYM theory are similar. Local vortex
operators exist, but the particles that carry vorticity are heavy in the continuum limit.

These results are broadly compatible with suggestions made in the
literature that at low values of the Chern Simons coefficient the YMCS
theory might undergo a phase transition. If this happens it is very
likely that this other
phase has a broken magnetic symmetry and is therefore confining.
This is a very interesting possibility which seems worthwhile
exploring by numerical lattice methods.

\section{Acknowledgments}
G.D. is supported in part by the U.S. DOE grant DE-FG02-92ER40716.00,
by PPARC grant PPA/V/S/1998/00910, and thanks Balliol College 
and the Theoretical Physics Department at Oxford for their hospitality.
A.K. is supported by PPARC.
The research of  B.\ T.\ is supported by  PPARC Grant PPA/G/O/1998/00567.

\vskip 1cm

\leftline{\bf References}  

\renewenvironment{thebibliography}[1]
        {\begin{list}{[$\,$\arabic{enumi}$\,$]}  
        {\usecounter{enumi}\setlength{\parsep}{0pt}
         \setlength{\itemsep}{0pt}  \renewcommand{\baselinestretch}{1.2}
         \settowidth
        {\labelwidth}{#1 ~ ~}\sloppy}}{\end{list}}

\myend